\documentclass[12pt,english,american,osajnl,preprint,showpacs,endfloats*,footinbib]{revtex4}
\usepackage{subfigure}
\usepackage{graphicx}

\makeatletter

\DeclareRobustCommand{\baselinestretch{2}}

\let\OrigMaketitle\maketitle
\renewcommand{\maketitle}{}

\makeatletter
\floats@sw{}%
{\setlength{\abovecaptionskip}{10mm}}
\renewcommand{\l@figure}[2]{\figurename\ #1\\[2cm]}
\makeatother

\let\OrigCaption\caption
\renewcommand{\caption}[2][X]{\OrigCaption[#2]{M. Xu et. al.}}

\usepackage{bm}
\def\P{\mathop{\operator@font P}\nolimits}

\usepackage{babel}
\makeatother
\begin{document}

\title{Multiple passages of light through an absorption inhomogeneity in
optical imaging of turbid media}

\author{M. Xu, W. Cai and R. R. Alfano}

\maketitle
\address{\small Institute for Ultrafast Spectroscopy and Lasers,\\ New York State Center
of Advanced Technology for Ultrafast Photonic Materials and Applications,\\ and Department of Physics,\\ The City College and Graduate Center of City University of New York, New York, NY 10031}

\email{minxu@sci.ccny.cuny.edu}

~

\begin{abstract}
The multiple passages of light through an absorption inhomogeneity
of finite size deep within a turbid medium is analyzed for optical
imaging using the {}``self-energy'' diagram. The nonlinear correction
becomes more important for an inhomogeneity of a larger size and with
greater contrast in absorption with respect to the host background.
The nonlinear correction factor agrees well with that from Monte Carlo
simulations for CW light. The correction is about $50\%-75\%$ in
near infrared for an absorption inhomogeneity with the typical optical
properties found in tissues and of size of five times the transport
mean free path. 
\end{abstract}

\OrigMaketitle\newpage{}

The main objective of optical imaging of turbid media is to locate
and identify the embedded inhomogeneities by essentially inverting
the difference in photon transmittance in time or frequency domains
due to the presence of these inhomogeneities.\cite{yodh95:_spect,arridge99:_optic,gandjbakhche.ea98:_time-dependent,cai.ea99:_optic}
The key quantity involved is the Jacobian which quantifies the influence
on the detected signal due to the change of the optical parameters
of the medium. The linear perturbation approach is only suitable to
calculate the Jacobian for a small and weak absorption inhomogeneity,
and not valid when the absorption strength is large.\cite{carraresi.ea2001:_accuracy}
This failure can be attributed to the multiple passages through the
abnormal site by the photon. The most important correction is the
{}``self-energy'' correction,\cite{negele98:_quant_many_system}
which takes into account the repeated visits made by a photon through
the site up to an infinite number of times. The presence of other
inhomogeneity {}``islands'' can be ignored because the photon propagator
decreases rapidly with the distance between two separate sites.

In this Letter, the nonlinear correction for an absorption inhomogeneity
of a large strength due to repeated visits by the photon is modeled
by a nonlinear correction factor (NCF) to the linear perturbation
approach. NCF as a function of the size and the strength of the inhomogeneity
is estimated using the {}``self-energy'' diagram. NCF is obtained
based on the cumulant approximation to radiative transfer, and verified
by Monte Carlo simulations for CW light. The magnitude of NCF is $0.5-1$
for an absorptive inhomogeneity of size up to $5l_{t}$ ($l_{t}$
is the mean transport free path of light), and of the typical optical
properties of human tissues ($\mu_{a}l_{t}/c\sim0.01-0.05$ where
$\mu_{a}$ is the absorption coefficient and $c$ is the speed of
light in the medium).

Consider an absorption site centered at $\bar{\mathbf{r}}$ and far
away from both the source and the detector, the change of the detected
light, $\Delta I$, at the detector $\mathbf{r}_{d}$ from a modulated
point source at $\mathbf{r}_{s}$ including the multiple passages
through the site, is given by\begin{eqnarray}
\Delta I & = & -G(\mathbf{r}_{d},\omega|\bar{\mathbf{r}})V\delta\mu_{a}(\bar{\mathbf{r}})\sum_{n=0}^{\infty}\left[-\bar{N}_{\mathrm{self}}(\omega;R)V\delta\mu_{a}(\bar{\mathbf{r}})\right]^{n}G(\bar{\mathbf{r}},\omega|\mathbf{r}_{s})\label{eq:DeltaI-nonlinear}\\
 & = & -G(\mathbf{r}_{d},\omega|\bar{\mathbf{r}})\frac{V\delta\mu_{a}(\bar{\mathbf{r}})}{1+\bar{N}_{\mathrm{self}}(\omega;R)V\delta\mu_{a}(\mathbf{\bar{\mathbf{r}}})}G(\bar{\mathbf{r}},\omega|\mathbf{r}_{s})\nonumber \end{eqnarray}
 where $\delta\mu_{a}$ is the excess absorption of the absorption
site of size $R$ and volume $V$, $\omega$ is the modulation frequency
of light, $G$ is the propagator of photon migration in the background
medium, and\begin{equation}
\bar{N}_{\mathrm{self}}(\omega;R)=\frac{1}{V^{2}}\int_{V}\int_{V}G(\mathbf{r}_{2},\omega|\mathbf{r}_{1})d^{3}\mathbf{r}_{2}d^{3}\mathbf{r}_{1}\label{eq:Nself-definition}\end{equation}
is the self-propagator which describes the probability that a photon
revisits the volume $V$. Here $G(\mathbf{r}_{2},\omega|\mathbf{r}_{1})$
gives the probability density that a photon \foreignlanguage{english}{leaves
the volume at $\mathbf{r}_{1}$ and reenters it at $\mathbf{r}_{2}$}.
\foreignlanguage{english}{The scattering property of the site is the
same as that of the background.} In Eq.~(\ref{eq:DeltaI-nonlinear})
$G(\mathbf{r}_{d},\omega|\bar{\mathbf{r}})$ and $G(\bar{\mathbf{r}},\omega|\mathbf{r}_{s})$
are well modeled by the center-moved diffusion model as long as the
separations $\left|\mathbf{r}_{d}-\bar{\mathbf{r}}\right|,\left|\mathbf{r}_{s}-\bar{\mathbf{r}}\right|\gg l_{t}$.\cite{xu01:_photon}
However, the diffusion Green's function can not be used in Eq.~(\ref{eq:Nself-definition})
to evaluate $\bar{N}_{\mathrm{self}}(\omega;R)$ because the diffusion
approximation breaks down when $\mathbf{r}_{1}$ is in the proximity
of $\mathbf{r}_{2}$. 

Comparing Eq.~(\ref{eq:DeltaI-nonlinear}) to the standard linear
perturbation approach, the nonlinear multiple passage effect of an
absorption site is represented by a nonlinear correction factor (NCF)\begin{equation}
\mathrm{NCF=}\left[1+\bar{N}_{\mathrm{self}}(\omega;R)V\delta\mu_{a}(\mathbf{\bar{\mathbf{r}}})\right]^{-1}.\label{eq:NCF}\end{equation}
This factor serves as a universal measure of the nonlinear multiple
passage effect as long as the absorption site is far away from both
the source and the detector and its size is much smaller than its
distance to both the source and the detector. This correction is more
significant when $\mathrm{NCF}$ is further away from unity.

The photon propagator $N(\mathbf{r}_{2},t|\mathbf{r}_{1},\mathbf{s})$,
the probability that a photon propagates from position $\mathbf{r}_{1}$
with propagation direction $\mathbf{s}$ to position $\mathbf{r}_{2}$
in time $t$, for any separation between $\mathbf{r}_{1}$ and $\mathbf{r}_{2}$,
was recently derived\cite{cai00:_cumul_boltz,xu01:_photon} in a form
of the cumulant approximation to radiative transfer. 

\begin{figure}
\begin{center}\subfigure[]{\includegraphics[%
  scale=0.9]{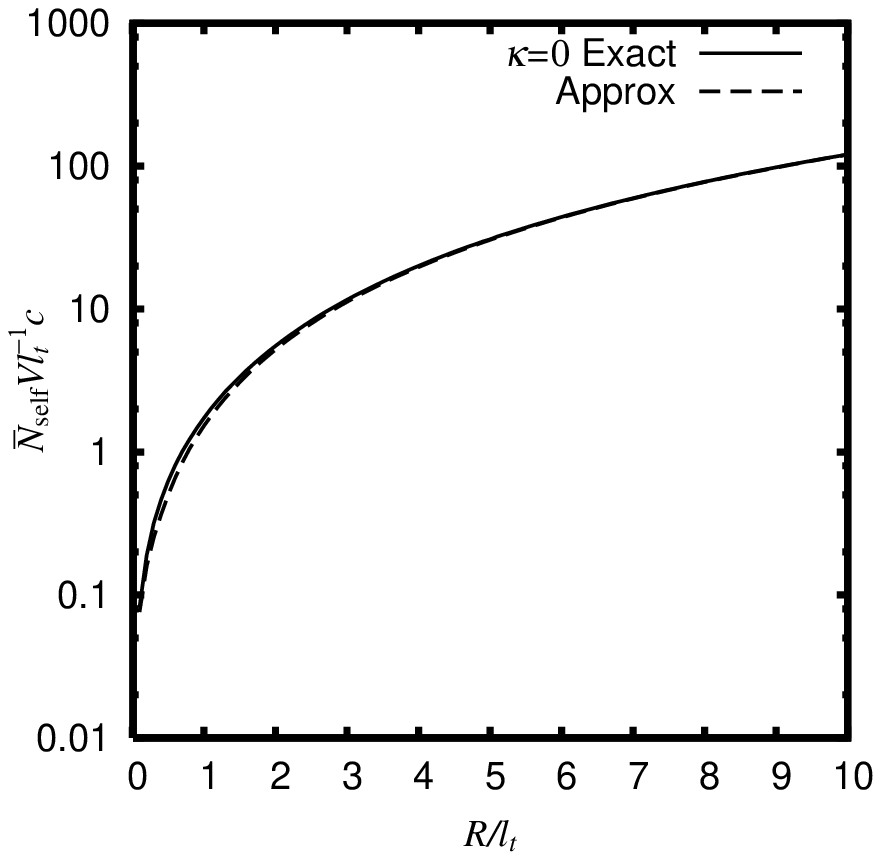}}~~\subfigure[]{\includegraphics[%
  scale=0.9]{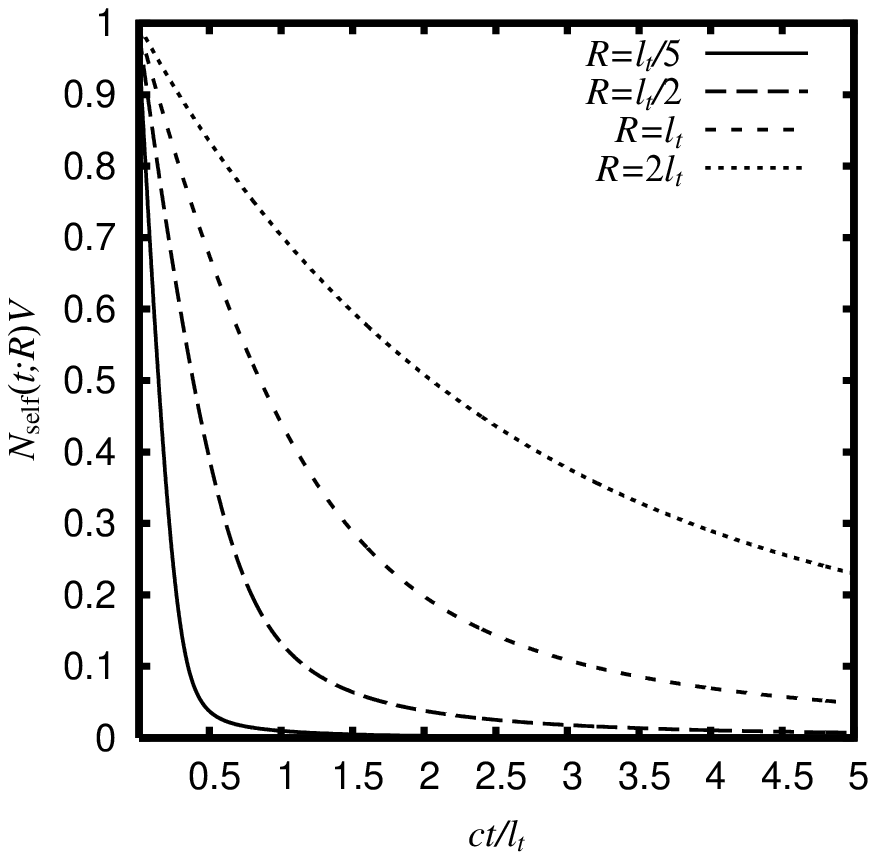}}\end{center}

\caption{(a) The self propagator $\bar{N}_{\mathrm{self}}(\omega;R)Vl_{t}^{-1}c$
and its approximation form when $\kappa=0$. (b) The self-propagator
for spheres of various radii in the time domain inside a nonabsorbing
medium. }

\label{cap:self-propagator-estimator}
\end{figure}

In the case of interest where the absorption site is deep inside the
medium, the photon distribution is isotropic. The photon propagator
is simplified to $N_{\mathrm{eff}}(r,t)\equiv N_{\mathrm{eff}}(\left|\mathbf{r}_{2}-\mathbf{r}_{1}\right|,t)$,
obtained by averaging $N(\mathbf{r}_{2},t|\mathbf{r}_{1},\mathbf{s})$
over the propagation direction $\mathbf{s}$ of light over the $4\pi$
solid angle. In the frequency domain, this effective propagator is
approximately given by\begin{equation}
N_{\mathrm{eff}}(r,\omega)\simeq\left\{ \begin{array}{cc}
\frac{1}{4\pi r^{2}c}\exp\left(-\frac{1}{3}\kappa^{2}l_{t}r\right)+\frac{\exp(-\kappa l_{t})}{4\pi Dr\kappa l_{t}}\sinh(\kappa r) & r<l_{t}\\
\frac{\exp\left(-\kappa r\right)}{4\pi Dr\kappa l_{t}}\sinh\left(\kappa l_{t}\right) & r\ge l_{t}\end{array}\right.\label{eq:Neff-asym-fourier}\end{equation}
 where $D\equiv l_{t}c/3$ and $\kappa\equiv\left[3(\mu_{a}-i\omega)/l_{t}c\right]^{1/2}$
whose sign is chosen with a nonnegative real part. The two terms in
$N_{\mathrm{eff}}$ when $r<l_{t}$ represent ballistic and diffusion
contributions respectively. The ballistic term does not depend on
scattering because the photon distribution involved is already isotropic.
Only diffusion contributes to $N_{\mathrm{eff}}$ when $r>l_{t}$.
The self propagator for an absorption sphere deep inside the medium
is given by:\begin{eqnarray}
\bar{N}_{\mathrm{self}}(\omega;R) & = & \frac{1}{V^{2}}\int_{V}\int_{V}N_{\mathrm{eff}}(\left|\mathbf{r}_{2}-\mathbf{r}_{1}\right|,\omega)d^{3}\mathbf{r}_{2}d^{3}\mathbf{r}_{1}\nonumber \\
 & = & \frac{1}{V}\int_{0}^{2R}N_{\mathrm{eff}}(r,\omega)\gamma_{0}(r)4\pi r^{2}dr\label{eq:Nself-fourier}\end{eqnarray}
where $\gamma_{0}(r)=1-\frac{3r}{4R}+\frac{1}{16}\left(\frac{r}{R}\right)^{3}$
is the characteristic function for a uniform sphere\cite{guinier55:_small_x}.
An absorption site of an arbitrary shape can be treated in the same
fashion. The exact self propagator must be computed by a numerical
quadrature. A good approximation of $\bar{N}_{\mathrm{self}}(\omega;R)$
is \begin{equation}
\bar{N}_{\mathrm{self}}(\omega;R)=\frac{l_{t}}{Vc}\left\{ \begin{array}{cc}
\left(\frac{3}{4}\xi+\xi^{3}\right)-\xi^{3}\kappa l_{t}+\mathcal{O}(\kappa^{2}) & \xi\le1/2\\
\left(\frac{6}{5}\xi^{2}+\frac{1}{2}-\frac{3}{16}\xi^{-1}+\frac{3}{320}\xi^{-3}\right)-\xi^{3}\kappa l_{t}+\mathcal{O}(\kappa^{2}) & \xi>1/2\end{array}\right.\label{eq:Nself-fourier-app}\end{equation}
 using Eq.~(\ref{eq:Neff-asym-fourier}) where $\xi\equiv R/l_{t}$
when $\left|\kappa\right|R\ll1$. This approximation (\ref{eq:Nself-fourier-app})
compares favorably to the exact $\bar{N}_{\mathrm{self}}(\omega=0;R)$
obtained by numerical quadrature for continuous wave light propagating
inside a nonabsorbing medium. The exact and approximate versions of
the dimensionless self-propagator $\bar{N}_{\mathrm{self}}Vl_{t}^{-1}c$
are plotted as solid and dashed lines, respectively, in Fig.~(\ref{cap:self-propagator-estimator}a).
The dimensionless self-propagator $\bar{N}_{\mathrm{self}}Vl_{t}^{-1}c$
depends solely on two dimensionless quantities $\kappa l_{t}$ of
the background and $R/l_{t}$ of the absorbing sphere.

It is worthwhile to point out that the self-propagator in time $\bar{N}_{\mathrm{self}}(t;R)$,
the inverse Fourier transform of (\ref{eq:Nself-fourier}), includes
the contribution from the ballistic motion of the photon when the
photon passes through the site. This ballistic contribution manifests
itself as the linear decay of $N_{\mathrm{seff}}(t;R)V$ in the form
of $\gamma_{0}(ct)$ near the origin of time, followed by a transition
to diffusion {[}Fig.~(\ref{cap:self-propagator-estimator}b){]}. 

The nonlinear correction factor is obtained by plugging (\ref{eq:Nself-fourier})
or (\ref{eq:Nself-fourier-app}) into (\ref{eq:NCF}). In particular,
we have \begin{equation}
\mathrm{NCF}\simeq\left\{ \begin{array}{cc}
\left[1+\frac{9}{16\pi}q\left(\xi^{-2}+\frac{4}{3}\right)\right]^{-1} & \xi\le1/2\\
\left[1+\frac{9}{10\pi}q\left(\xi^{-1}+\frac{5}{12}\xi^{-3}-\frac{5}{32}\xi^{-4}+\frac{1}{128}\xi^{-6}\right)\right]^{-1} & \xi>1/2\end{array}\right.\label{eq:NCF-app}\end{equation}
where $q\equiv V\delta\mu_{a}(\mathbf{\bar{\mathbf{r}}})/l_{t}^{2}c$
is the dimensionless strength of the absorber when $\left|\kappa\right|R\ll1$.
The effectiveness of an absorber of a fixed strength depends on its
size. For an absorber of a fixed $q>0$, the effectiveness of absorbing
light is diminished ($\mathrm{NCF}$ decreases) when its size is reduced.
This can be understood from the fact that the photon spends less time
per volume inside the absorber of a smaller dimension because of the
ballistic motion of the photon after each scattering event. The photon
leaves a small site $(R<l_{t}$) in an almost straight line. The diffusion
behavior for an individual photon is only observed after a large number
of scattering and on a scale larger than $l_{t}$. 

\begin{figure}
\begin{center}\includegraphics[%
  scale=0.9]{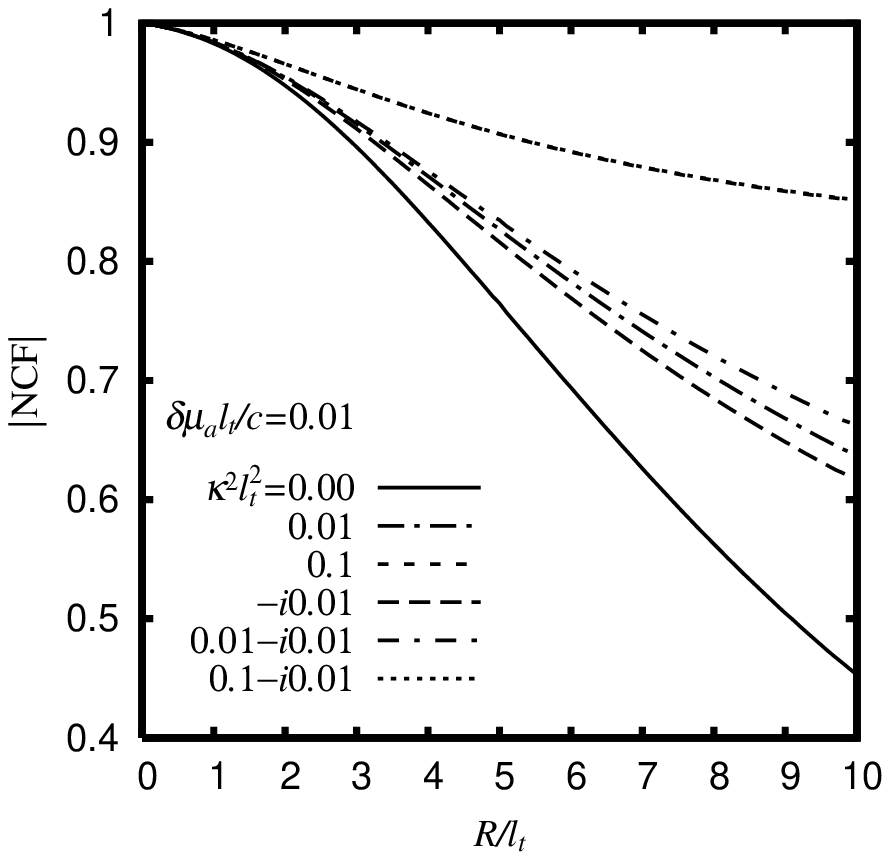}~~\includegraphics[%
  scale=0.9]{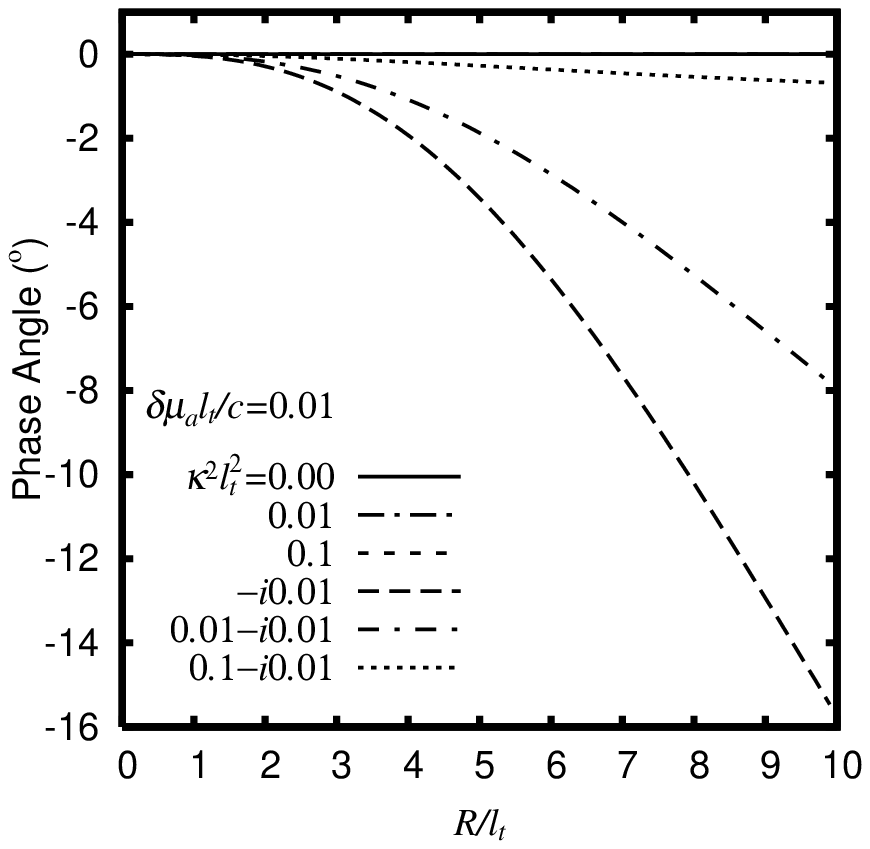}\end{center}

\begin{center}\includegraphics[%
  scale=0.9]{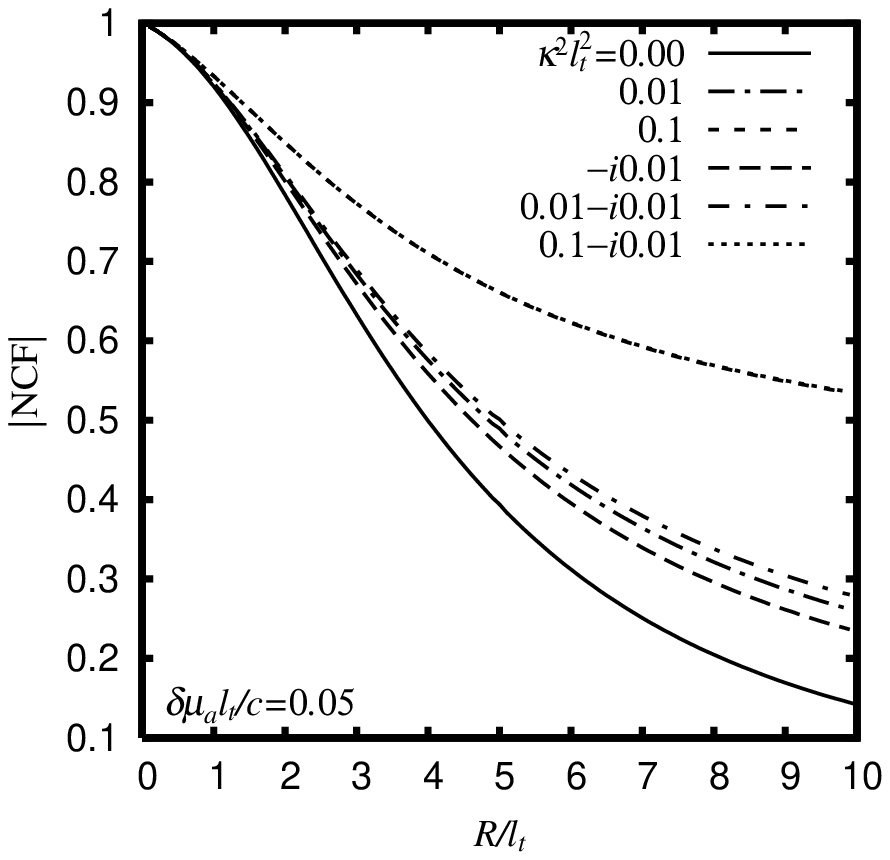}~~\includegraphics[%
  scale=0.9]{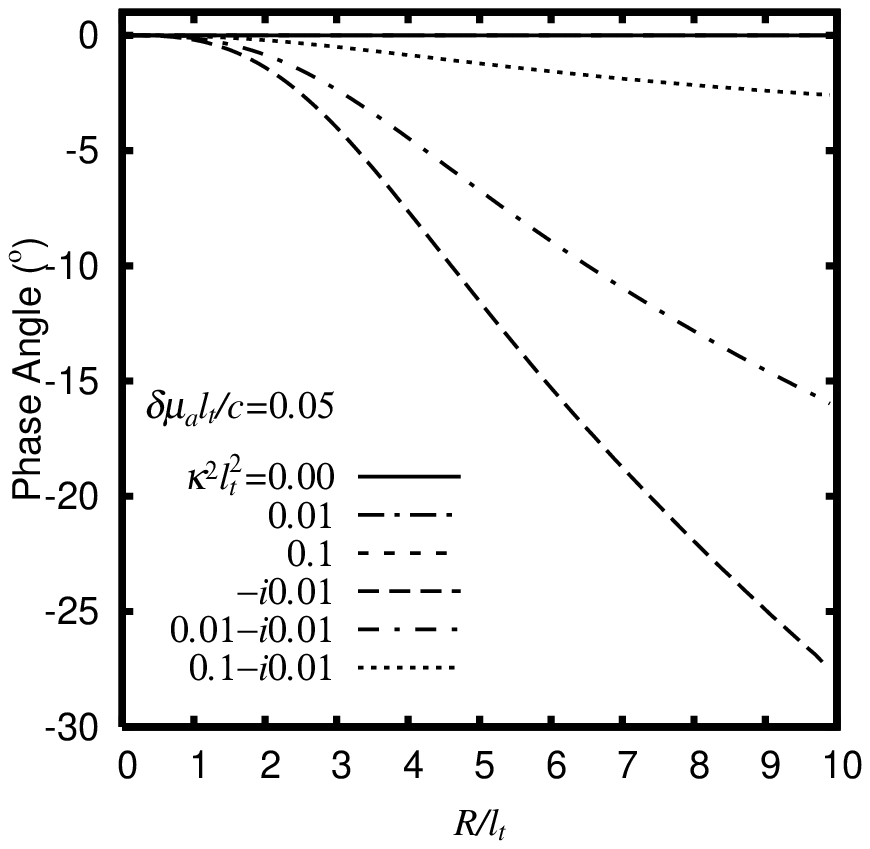}\end{center}

\caption{The nonlinear correction factor (magnitude and phase angle) vs the
size of absorbers whose excess absorption $\delta\mu_{a}l_{t}/c$
equals $0.01$ and $0.05$, respectively. Note $\kappa^{2}l_{t}^{2}=3(\mu_{a}-i\omega)l_{t}/c$
for the background medium.}

\label{cap:NCF}
\end{figure}

Fig.~(\ref{cap:NCF}) shows plots of NCF versus absorber size for
typical absorbers of excess absorption $\delta\mu_{a}l_{t}/c$ equal
$0.01$ and $0.05$, respectively. The nonlinear correction factor
generally decreases with the size of the absorber whose excess absorption
is fixed. With the increase of the background absorption and the modulation
frequency, the nonlinear correction becomes less accentuated. The
phase delay is larger for higher modulation frequencies and less background
absorption. 

\begin{figure}
\begin{center}\subfigure[]{\includegraphics[%
  scale=0.9]{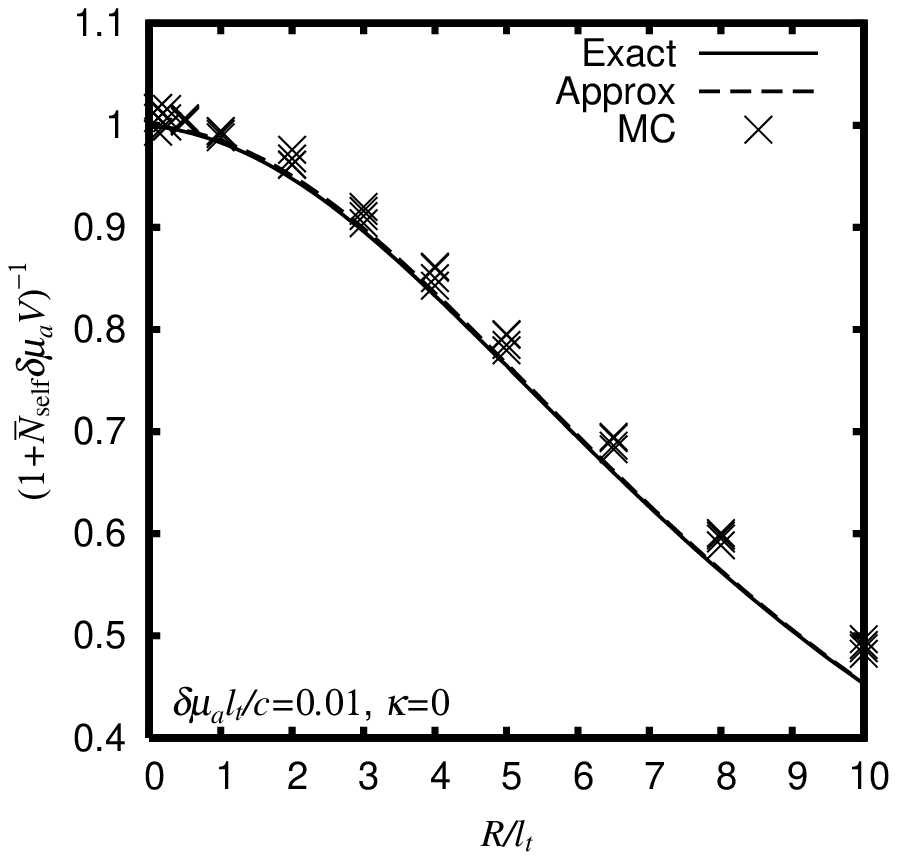}}~~\subfigure[]{\includegraphics[%
  scale=0.9]{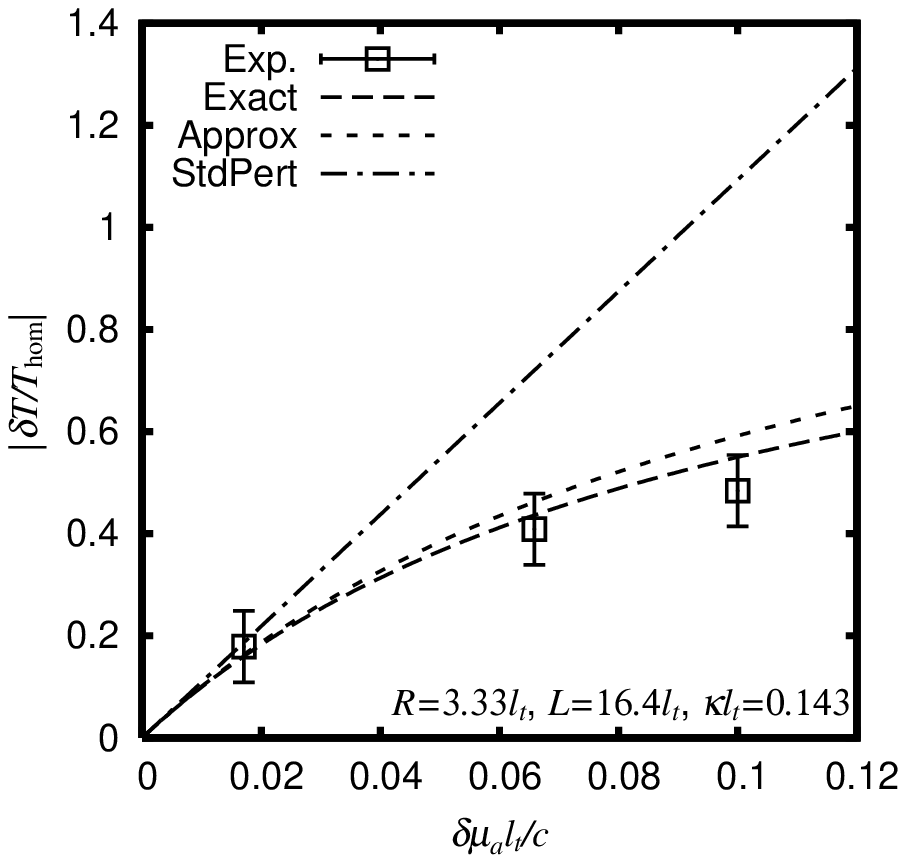}}\end{center}

\caption{(a) The theoretical nonlinear correction factors from numerical quadrature
(Exact), the approximate form Eq.~(\ref{eq:NCF-app}) (Approx), and
Monte Carlo simulations (MC). \foreignlanguage{english}{Results from
four independent MC simulations are shown for each} radius. The standard
linear perturbation approach corresponds to a horizontal line $\mathrm{NCF}=1$
(not shown in the figure). (b) The percentage change of the CW transmittance
from the experimental data given in Fig.~(9) of Ref.~(\onlinecite{carraresi.ea2001:_accuracy})
compared to the theoretical predictions made by the standard linear
perturbation approach (StdPert) and the ones including NCF (Exact
and Approx).}

\label{cap:NCF-MC}
\end{figure}

Monte Carlo simulations\cite{testorf.ea99:_sampling} are performed
for CW light propagating in a uniform nonabsorbing and isotropic scattering
slab. The thickness of the slab is $L=80l_{t}$. A spherical absorber
of radius $R$ is located at the center $(0,0,L/2)$ of the slab.
The excess absorption of the absorber is $\delta\mu_{a}l_{t}/c=0.01$.
The absorber has the same scattering property as the background. The
details of the MC computation has been provided in a previous publication\cite{xu01:_photon_migrat}.
The correlated sampling method is used in each simulation to reduce
variance.\cite{rief94:_monte_carlo} A single simulation is used to
compute the change in light transmittance due to the presence of the
absorption site and the corresponding nonlinear correction factor.
Fig.~(\ref{cap:NCF-MC}a) shows the nonlinear correction factors
obtained from numerical quadrature, the approximate form Eq.~(\ref{eq:NCF-app}),
and Monte Carlo simulations, respectively. The agreement between our
theoretical NCF and that from Monte Carlo simulations is excellent.
The slight difference between them at large radii is accounted for
by the fact that the sphere can no longer be regarded as small compared
to the dimensions of the slab. The probability of a photon revisiting
a large sphere is overestimated by Eq.~(\ref{eq:Nself-fourier})
for the sphere located at the center of the slab.%
\footnote{Consider the facts: (1) the probability of a photon revisiting the
sphere decreases when the position from which the photon leaves the
sphere is further away from the center of the sphere, and (2) the
photon density inside the sphere is higher in regions closer to its
surface for the sphere located at the center of the slab. The arithmetic
mean taken in Eq.~(\ref{eq:Nself-fourier}) hence overestimates the
revisiting probability.%
} 

Fig.~(\ref{cap:NCF-MC}b) shows the percentage change of the CW transmittance
estimated from the experimental data given in Fig.~(9) of Ref.~(\onlinecite{carraresi.ea2001:_accuracy}).
The relevant parameters of the experiment are summarized in the inset.
The theoretical predictions from the linear perturbation approach
with and without the nonlinear correction are also shown in Fig.~(\ref{cap:NCF-MC}b)
assuming a collimated point source and a point detector in a cofocal
setup. The agreement between our theoretical prediction with nonlinear
correction and the experimental result is good.

The typical value of the absorption coefficient of human tissues in
the near infrared indicates that $\mu_{a}l_{t}/c\sim0.01-0.05$.\cite{peters.ea90:_optic,cheong.ea90:_revie}
This fact should put our results on NCF in this range {[}Figs.~(\ref{cap:NCF})
and (\ref{cap:NCF-MC}){]} into perspective. Our analysis reveals
that the nonlinear correction for absorption inhomogeneities due to
multiple passages of light in optical imaging of human tissues becomes
important when the size of the inhomogeneity is much larger than the
mean transport free path of light. This factor is smaller (further
away from unity) for higher excess absorption and weaker background
absorption. The value of $\mathrm{NCF}$ decreases from $\sim0.75$
to $\sim0.5$ for an absorption site of radius $5l_{t}$ with excess
absorption $\delta\mu_{a}l_{t}/c$ increasing from $0.01$ to $0.05$.
This corresponds to $75\%$ and $50\%$ underestimations of the excess
absorption coefficient in these cases if the standard linear perturbation
approach is applied naively. 

In conclusion, the nonlinear correction is analyzed for the effect
of the multiple passages through an absorption inhomogeneity of finite
size deep inside a turbid medium in optical imaging using the {}``self-energy''
diagram. The nonlinear correction factor is verified by Monte Carlo
simulations for CW light and supported by experimental results. The
nonlinear correction becomes more important for an inhomogeneity of
a larger size and with greater contrast in absorption with respect
to the background. The standard linear perturbation approach in optical
imaging should be augmented to include this nonlinear correction.

This work is supported in part by NASA and Army. M. X. thanks the
support by the Department of Army (Grant\# DAMD17-02-1-0516). The
U. S. Army Medical Research Acquisition Activity, 820 Chandler Street,
Fort Detrick MD 21702-5014 is the awarding and administering acquisition
office. The authors are indebted to the anonymous referees who help
in improving this paper.

\printtables \clearpage \listoffigures \clearpage \printfigures

\begin{thebibliography}{10}

\bibitem{yodh95:_spect}
A.~Yodh and B.~Chance, Phys. Today {\bf 48}, 38 (1995).

\bibitem{arridge99:_optic}
S.~R. Arridge, Inverse Problems {\bf 15}, R41 (1999).

\bibitem{gandjbakhche.ea98:_time-dependent}
A.~H. Gandjbakhche, V.~Chernomordik, J.~C. Hebden and R.~Nossal, Appl. Opt.
  {\bf 37}, 1973 (Apr. 1998).

\bibitem{cai.ea99:_optic}
W.~Cai, S.~K. Gayen, M.~Xu, M.~Zevallos, M.~Alrubaiee, M.~Lax and R.~R. Alfano,
  Appl. Opt. {\bf 38}, 4237 (1999).

\bibitem{carraresi.ea2001:_accuracy}
S.~Carraresi, T.~S.~M. Shatir, F.~Martelli and G.~Zaccanti, Appl. Opt. {\bf
  40}, 4622 (Sep. 2001).

\bibitem{negele98:_quant_many_system}
J.~W. Negele and H.~Orland, {\em Quantum Many-particle Systems\/}, (Westview
  Press, Boulder, Colorado) (1998).

\bibitem{xu01:_photon}
M.~Xu, W.~Cai, M.~Lax and R.~R. Alfano, Opt. Lett. {\bf 26}, 1066 (2001).

\bibitem{cai00:_cumul_boltz}
W.~Cai, M.~Lax and R.~R. Alfano, Phys. Rev. E {\bf 61}, 3871 (2000).

\bibitem{guinier55:_small_x}
A.~Guinier, G.~Fournet, C.~B. Walker and K.~L.Yudowitch, {\em Small-angle
  scattering of X-rays\/}, (John Wiley \& Sons, New York) (1955).

\bibitem{testorf.ea99:_sampling}
M.~Testorf, U.~Osterberg, B.~Pogue and K.~Paulsen, Appl. Opt. {\bf 38}, 236
  (January 1999).

\bibitem{xu01:_photon_migrat}
M.~Xu, W.~Cai, M.~Lax and R.~R. Alfano, Phys. Rev. E {\bf 65}, 066609 (2002).

\bibitem{rief94:_monte_carlo}
H.~Rief, J. Comput. Phys. {\bf 111}, 33 (1994).

\bibitem{peters.ea90:_optic}
V.~G. Peters, D.~R. Wyman, M.~S. Patterson and G.~L. Frank, Phys. Med. Biol.
  {\bf 35}, 1317 (1990).

\bibitem{cheong.ea90:_revie}
W.~F. Cheong, S.~Prahl and A.~J. Welch, IEEE J. Quant. Elecron. {\bf 26}, 2166
  (1990).

\end{thebibliography}
\end{document}